\documentclass[aps,10pt,a4paper,twocolumn,superscriptaddress]{revtex4-2}
\usepackage[utf8]{inputenc}
\usepackage{amsmath}
\usepackage{amssymb}
\usepackage{braket}
\usepackage{graphicx}
\usepackage{xcolor}
\usepackage{lipsum}
\usepackage{physics,mathtools}
\usepackage{siunitx}

\newcommand{\qtr}{\operatorname{Tr}}
\newcommand{\ctr}{{\operatorfont{\mathsf{Tr}}}}
\newcommand{\qtrabs}{\operatorname{TrAbs}}
\newcommand{\ctrabs}{{\operatorname{\mathsf{TrAbs}}}}

\renewcommand*{\det}{{\operatorname{\mathsf{Det}}}}

\newcommand{\re}{\operatorname{Re}}
\newcommand{\btheta}{{\boldsymbol{\theta}}}
\newcommand{\hbtheta}{\skew{2.5}\hat{\boldsymbol{\theta}}}
\newcommand{\F}{\mathcal{F}}

\newtheorem{thm}{Theorem}

\begin{document}

\title{Comparison of estimation limits for quantum two-parameter estimation}

\date{\today}
\author{Simon K. Yung}
\email{sksyung@gmail.com}
\affiliation{Centre for Quantum Computation and Communication Technology, Department of Quantum Science and Technology, Research School of Physics, The Australian National University, Canberra, ACT 2601, Australia.}
\affiliation{A*STAR Quantum Innovation Centre (Q.INC), Institute of Materials Research and Engineering (IMRE), Agency for Science, Technology and Research (A*STAR), 2 Fusionopolis Way, Innovis, 138634, Singapore.}
\author{Lorc\'an O. Conlon}
\email{lorcanconlon@gmail.com}
\affiliation{A*STAR Quantum Innovation Centre (Q.INC), Institute of Materials Research and Engineering (IMRE), Agency for Science, Technology and Research (A*STAR), 2 Fusionopolis Way, Innovis, 138634, Singapore.}
\author{Jie Zhao}
\affiliation{Centre for Quantum Computation and Communication Technology, Department of Quantum Science and Technology, Research School of Physics, The Australian National University, Canberra, ACT 2601, Australia.}
\author{Ping Koy Lam}
\affiliation{A*STAR Quantum Innovation Centre (Q.INC), Institute of Materials Research and Engineering (IMRE), Agency for Science, Technology and Research (A*STAR), 2 Fusionopolis Way, Innovis, 138634, Singapore.}
\affiliation{Centre for Quantum Computation and Communication Technology, Department of Quantum Science and Technology, Research School of Physics, The Australian National University, Canberra, ACT 2601, Australia.}
\affiliation{Centre for Quantum Technologies, National University of Singapore, 3 Science Drive 2, Singapore 117543, Singapore.}
\author{Syed M. Assad}
\email{cqtsma@gmail.com}
\affiliation{A*STAR Quantum Innovation Centre (Q.INC), Institute of Materials Research and Engineering (IMRE), Agency for Science, Technology and Research (A*STAR), 2 Fusionopolis Way, Innovis, 138634, Singapore.}
\affiliation{Centre for Quantum Computation and Communication Technology, Department of Quantum Science and Technology, Research School of Physics, The Australian National University, Canberra, ACT 2601, Australia.}

\begin{abstract}
    Measurement estimation bounds for local quantum multiparameter estimation, which provide lower bounds on possible measurement uncertainties, have so far been formulated in two ways: by extending the classical Cram\'{e}r--Rao bound (e.g., the quantum Cram\'{e}r--Rao bound and the Nagaoka Cram\'er--Rao bound) and by incorporating the parameter estimation framework with the uncertainty principle, as in the Lu--Wang uncertainty relation. In this work, we present a general framework that allows a direct comparison between these different types of estimation limits. Specifically, we compare the attainability of the Nagaoka Cram\'er--Rao bound and the Lu--Wang uncertainty relation, using analytical and numerical techniques. We show that these two limits can provide different information about the physically attainable precision. We present an example where both limits provide the same attainable precision and an example where the Lu--Wang uncertainty relation is not attainable even for pure states. We further demonstrate that the unattainability in the latter case arises because the figure of merit underpinning the Lu--Wang uncertainty relation (the difference between the quantum and classical Fisher information matrices) does not necessarily agree with the conventionally used figure of merit (mean squared error). The results offer insights into the general attainability and applicability of the Lu--Wang uncertainty relation. Furthermore, our proposed framework for comparing bounds of different types may prove useful in other settings. 
\end{abstract}

\maketitle

Parameter estimation is important for optimising experimental processes and efficiently extracting information about physical systems. The parameter estimation theory of quantum systems has been developed from classical theory, with analogues of Fisher information and the Cram\'er--Rao bound remaining prominent \cite{fisher_mathematical_1922,harald_cramer_mathematical_1946,rao_minimum_1947,rao_information_1992,helstrom_minimum_1967,helstrom_minimum_1968,braunstein_statistical_1994,helstrom_quantum_1976,paris_quantum_2009}. 

\begin{figure*}
    \centering
    \includegraphics{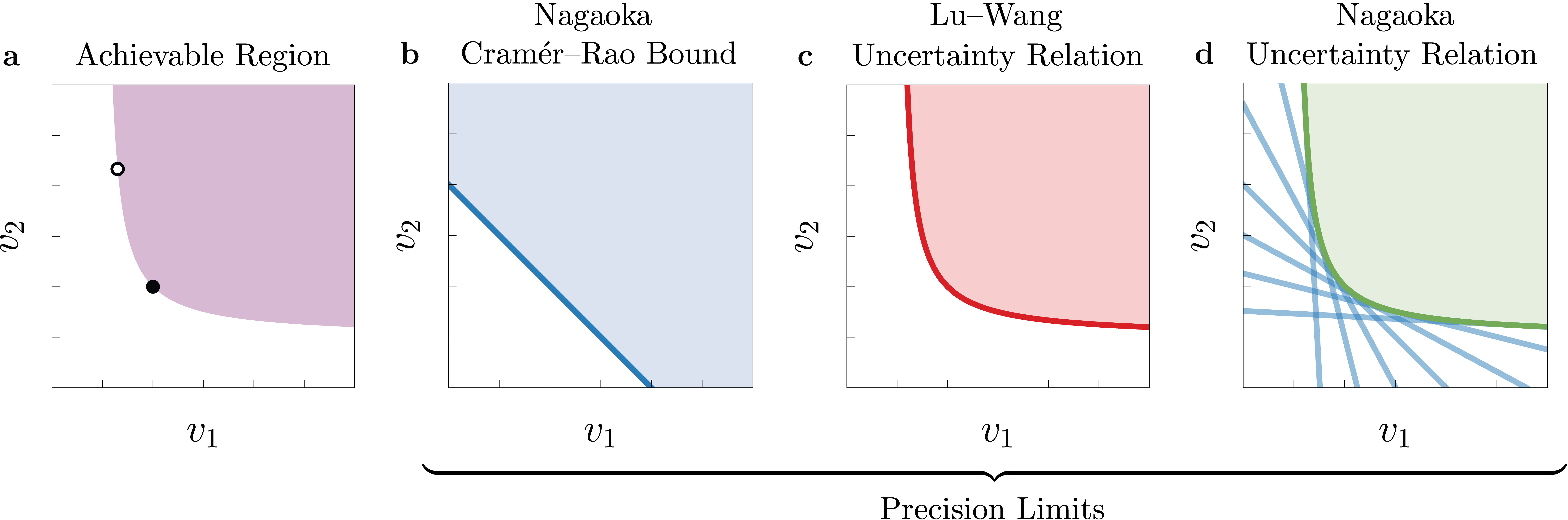}
    \caption{Graphical depiction of achievable estimation errors and estimation limits. Here, $v_i$ denotes the mean squared error of the estimate of the parameter $\theta_i$, in arbitrary units. \textbf{a}, in general, there is a trade-off between optimising each parameter's estimate. The shaded region illustrates pairs $(v_1,v_2)$ that are possible. The solid marker denotes the errors of a measurement that minimises the total error, while the open marker denotes the errors of a measurement that places more importance on minimising the error of the first parameter. \textbf{b}, the Nagaoka Cram\'er--Rao bound (and other Cram\'er--Rao type bounds) restricts a region bounded by a straight line. The weighting dictates the gradient of the line. \textbf{c}, the Lu--Wang uncertainty relation defines a forbidden region that is bounded by a curve. \textbf{d} by combining the regions for different weightings of the Nagaoka Cram\'er--Rao bound, an overall region (green) can be determined, defining the Nagaoka uncertainty relation.}     \label{fig:weightexample}
\end{figure*}

For the simultaneous estimation of multiple parameters of a quantum system, complications stem from the incompatibility of operators and the uncertainty principle \cite{liu_quantum_2019,albarelli_perspective_2020,demkowicz-dobrzanski_multi-parameter_2020,sidhu_geometric_2020}. Notably, in contrast to its classical counterpart, quantum multiparameter estimation is subject to trade-offs between minimising the mean squared error (MSE) for each parameter \cite{baumgratz_quantum_2016}, see Figure \ref{fig:weightexample}\textbf{a}. This complicates determining the optimal measurement process, as all parameters must be considered simultaneously. Rather than resorting to brute force optimisation to determine the best measurement and its precision, it is typical to use limits on the precision that are easier to compute to guide analysis.  

In this work, we investigate two precision limits for two-parameter estimation that were developed in different ways and a priori have different advantages. The first limit is the Nagaoka Cram\'er--Rao bound (NCRB), an extension of the classical estimation limits \cite{nagaoka_new_2005}. It places a lower bound on the trace of the MSE matrix, or, notionally, the total error, as illustrated in Figure \ref{fig:weightexample}\textbf{b}. In general, the computation of the NCRB requires the solving of a minimisation problem. While the general attainability of the NCRB is unknown, it is known to be attainable for estimation in qubit systems and is thus a useful benchmark \cite{nagaoka_new_2005,nagaoka_generalization_2005}. The second limit is the Lu--Wang uncertainty relation (LWUR), which is a non-trivial trade-off relation for mean squared errors~\cite{lu_incorporating_2021}, see Figure \ref{fig:weightexample}\textbf{c}. As such, it indicates the trade-off between optimising each parameter's estimate but does not immediately indicate the total error. The LWUR was derived from an uncertainty relation for the uncertainties in the joint measurement of any two quantum operators, developed by Ozawa~\cite{ozawa_universally_2003,ozawa_uncertainty_2004,ozawa_error-disturbance_2014} and Branciard~\cite{branciard_error-tradeoff_2013}. An advantage of the LWUR is that minimisation is not required to compute the uncertainty relation. However, the general attainability of the LWUR has not been studied despite its application to several physical estimation problems.  

While both of the above-mentioned precision limits restrict the MSE, they do so in different ways, either limiting the trace of the MSE matrix or imposing a trade-off between different elements. As such, their comparison is not straightforward, and, to the best of our knowledge, there does not currently exist any comprehensive method for such a comparison. Partial methods do exist, such as the construction of uncertainty relations from Cram\'er--Rao bounds developed by Kull \textit{et al.~}\cite{kull_uncertainty_2020} and Funada and Suzuki \cite{funada_error_2020,funada_trade-off_2022,suzuki_parameter_2015} and illustrated in Figure \ref{fig:weightexample}\textbf{d}. We apply this method to the NCRB and further develop methods to complete the comparison to the LWUR.  With these methods we find scenarios where the LWUR is as informative as the NCRB, but also scenarios where it is less informative.  

\section{Background}
\subsection{Preliminaries}
We consider a quantum state described by a density operator $\rho_\btheta$, with two real parameters $\btheta = (\theta_1,\theta_2)^\top$. The aim is to estimate the value of $\btheta$ by implementing a measurement described by an $m$-outcome POVM $\Pi = \{\Pi_1,\dots,\Pi_m \ | \ \sum_{i=1}^m \Pi_i=\openone\}$. The measurement is repeated $N$ times (assumed to be large) and the results are processed to obtain the observed frequencies of the outcomes.  
The post-processing is performed using an estimator, $\hbtheta = (\hat{\theta}_1,\hat{\theta}_2)^\top$, which maps from the measurement outcomes to the parameter estimates. We consider local estimation with the prior knowledge that the true parameter value is approximately $\btheta \approx \btheta_0$. This assumption is well justified when a sufficiently large number of copies of the quantum state $\rho_\btheta$ are available, as a small portion can be used to determine $\btheta_0$. We wish to optimise for sensing small variations of the parameters near $\btheta_0$, and we assume that the estimators are locally unbiased, satisfying
\begin{equation}
    \mathbb{E}\left[ \hat{\theta}_i \right]\Big|_{\btheta= \btheta_0} = \theta_i, \quad \frac{\partial}{\partial \theta_j}\mathbb{E}\left[ \hat{\theta}_i\right]\Big|_{\btheta =\btheta_0} = \delta_{ij}.
\end{equation} 
The locally unbiased conditions ensure that the estimator is unbiased to first order around $\btheta_0$; these conditions are weaker than global unbiasedness but are sufficient to ensure accurate estimates can be obtained using sufficient resources. 

The performance of the estimation process is quantified by the estimator's mean squared error (MSE) matrix, $V(\hbtheta)$, with elements
\begin{equation}
    [V(\hbtheta)]_{jk} = \mathbb{E}\left[ \left( \hat{\theta}_j-\theta_j \right)\left( \hat{\theta}_k-\theta_k \right) \right], \ j,k \in \{1,2\}, \label{eq:mse}
\end{equation}
i.e., the expected squared deviation between the estimated and true values of the parameters. For a locally unbiased estimator, the MSE matrix coincides with its covariance matrix. The typical goal is to minimise its trace.

For a given POVM, the classical Fisher information matrix of the effective probability distribution has elements
\begin{equation}
    F_{ij} = \mathbb{E}_{k}\left[ \frac{\partial\log p(k|\btheta)}{\partial \theta_i}\frac{\partial\log p(k|\btheta)}{\partial \theta_j}  \right], \label{eq:fisher}
\end{equation}  
where $k$ labels the measurement outcomes. The Fisher information is useful because its inverse places a lower bound on the MSE of any unbiased estimator \cite{harald_cramer_mathematical_1946,rao_information_1992,rao_minimum_1947}, called the classical Cram\'er--Rao bound
\begin{equation}
    V(\hbtheta) \succeq \frac{1}{N}F^{-1},
\end{equation}
where $N$ is the number of samples. Hereafter, the factor $1/N$ will be suppressed, equivalent to considering the MSE per sample. The Cram\'er--Rao bound can be converted into a scalar lower bound by taking a (possibly weighted) trace of both sides of the inequality. For a given classical probability distribution, the classical Cram\'er--Rao bound can be saturated in the asymptotic limit using the maximum likelihood estimator \cite{fisher_theory_1925}. However, the lower bound is measurement-dependent and is thus a statement about the processing of results. Nevertheless, this property allows the classical Fisher information to be used to determine the possible MSE of a measurement. 

\subsection{Quantum Cram\'er--Rao Bound}
Measurement-independent analogues of the classical Fisher information provide a measure of the information that any measurement can extract. The most commonly-used version is based on the symmetric logarithmic derivative (SLD) operator, a generalisation of the logarithmic derivative in Eq.\ \eqref{eq:fisher} for a density operator. These operators, $L_i$, are defined implicitly by
\begin{equation}
    \frac{\partial \rho_\btheta}{\partial \theta_i} = \frac{1}{2}\left( \rho_\btheta L_i +L_i \rho_\btheta \right).
\end{equation}
The symmetric logarithmic derivative quantum Fisher information (SLDQFI) matrix, sometimes simply called the quantum Fisher information, has elements
\begin{equation}
    \mathcal{F}_{ij}= \re \qtr[\rho_\btheta L_i L_j]. \label{eq:sldqfi}
\end{equation}
The SLDQFI places a bound on the classical Fisher information of any measurement \cite{braunstein_statistical_1994}
\begin{equation}
    F \preceq \mathcal{F},
\end{equation}
and accordingly places a measurement-independent lower bound on the MSE, called the SLD Cram\'er--Rao bound (SLDCRB). This is also known as the Helstrom CRB. The SLDCRB is tight for single-parameter estimation and can be saturated by measuring in the SLD operator's eigenbasis. However, the SLD operators need not commute for systems with multiple parameters, so simultaneous optimal estimation of all parameters is not always possible. The optimal measurements for individually estimating each parameter may not be compatible, rendering the SLDCRB unattainable in general. 

In this work, we largely restrict our attention to separable measurements that are performed on individual copies of the probe state. Measurements on multiple copies simultaneously, called collective measurements, are also possible via entanglement. Due to the increased degrees of freedom, collective measurements can more efficiently extract information about the system, offering a uniquely quantum advantage \cite{chen_information_2022,chen_incompatibility_2022}. This improvement, however, comes at the cost of experimental complexity. At present, only few-copy collective measurements have been successfully demonstrated \cite{hou_deterministic_2018,conlon_approaching_2023,zhou_experimental_2023,conlon_discriminating_2023}. We focus on estimation limits for separable measurements, but use collective measurement bounds as a point of comparison.

\subsection{Tighter Cram\'er--Rao Bounds}

Several Cram\'er--Rao type lower bounds, i.e., those that place a bound on a weighted trace of the MSE matrix, have been developed and are tighter than the SLDCRB. Here, we focus on the Nagaoka Cram\'er--Rao bound (NCRB), which applies only to separable measurements~\cite{nagaoka_new_2005}. This lower bound is the solution of a minimisation problem over Hermitian matrices $X = (X_1,X_2)^\top$ satisfying locally unbiased conditions. It can be written as 
\begin{multline}
    \ctr[WV(\hbtheta)]\geq \mathcal{C}_{\text{N}} \coloneq \min_X \Big\{ \ctr[W Z_\btheta[X]] \\ + \sqrt{\det[W]}\qtrabs[\rho_\btheta[X_1,X_2]] \Big\}, \label{eq:nagfull}
\end{multline}
where $W$ is called the weight matrix and is positive-definite. The weight matrix defines the relative importance of each parameter. Here, $Z_\btheta[X]$ has elements \mbox{$Z_\btheta[X]_{ij} = \qtr[\rho_\btheta X_i X_j]$} and $\qtrabs[Y]$ is the sum of the absolute values of the eigenvalues of $Y$. For the estimation of parameters encoded in qubits, Nagaoka demonstrated that the NCRB is attainable and has the following analytic form~\cite{nagaoka_new_2005,nagaoka_generalization_2005}
\begin{equation}
    \mathcal{C}_{\text{N},d=2} = \ctr[W \mathcal{F}^{-1}] + 2\sqrt{\det[W\mathcal{F}^{-1}]}. \label{eq:NCRBanalytic}
\end{equation}
For higher-dimensional systems, the NCRB minimisation problems can be computed as a semidefinite program~\cite{conlon_efficient_2021}. The NCRB (for two-parameter estimation) was conjectured to be tight \cite{nagaoka_generalization_2005}. However, its extension to more parameters, called the Nagaoka--Hayashi Cram\'er--Rao bound \cite{nagaoka_generalization_2005,hayashi_simultaneous_1999,conlon_efficient_2021}, was recently demonstrated to not be tight in general \cite{hayashi_tight_2022}.

For collective measurements on any number of copies, the Holevo Cram\'er--Rao bound (HCRB) can be used instead. It is also the solution of a minimisation problem and can be written as
\begin{equation}
    \mathcal{C}_{\text{H}} \coloneq \min_X \left\{ \ctr[W \re Z_\btheta[X]]+\ctrabs[W \operatorname{Im}Z_\btheta[X]] \right\}.
\end{equation}
For pure states, the HCRB is equal to the NCRB \cite{matsumoto_new_2002}, but the HCRB cannot in general be attained using single-copy measurements---the ratio between the single-copy bound (the NCRB or Nagaoka--Hayashi CRB) and the HCRB can be as large as linear in the dimension of the Hilbert space \cite{das_holevo_2024}. However, the HCRB can be saturated by an asymptotic collective measurement \cite{kahn_local_2009,yamagata_quantum_2013,yang_attaining_2019}. On the other hand, if the HCRB is not saturated by separable measurements, it cannot be saturated by a collective measurement on any finite number of copies \cite{conlon_gap_2022}.  

The Cram\'er--Rao bounds have no explicit reference to Heisenberg's uncertainty relation. Instead, the uncertainty principle is built into the definition through the constraints of the minimisation problems. As such, the lower bounds alone do not indicate the balance of the elements of the MSE matrix (Eq.\ \eqref{eq:mse}) or the trade-off between them. In Sec.~\ref{sec:methods}, we demonstrate how the Cram\'er--Rao bounds can be extended to extract this information. 

\subsection{Uncertainty Relations for Parameter Estimation}

Lu and Wang introduced a new uncertainty relation for multiparameter estimation by defining the ``information regret'' matrix of a measurement as the difference between the classical and quantum Fisher information matrices, \mbox{$R = \mathcal{F}-F$}~\cite{lu_incorporating_2021}. Lu and Wang used Ozawa's ``universally valid uncertainty relation'' \cite{ozawa_universally_2003,ozawa_uncertainty_2004,branciard_error-tradeoff_2013,ozawa_error-disturbance_2014} to form a trade-off relation between the elements of the regret matrix. Called the Information Regret Trade-off Relation (IRTR), it is given by
\begin{equation}
    \Delta_1^2+\Delta_2^2+2\sqrt{1-\tilde{c}^2}\Delta_1\Delta_2 \geq \tilde{c}^2. \label{eq:IRTR}
\end{equation}
Here, $\Delta_j = \sqrt{R_{jj}/\mathcal{F}_{jj}}$ is the normalised-square-root regret. The incompatibility coefficient, $\tilde{c}$, which takes values $0\leq \tilde{c}\leq 1$, is defined by
\begin{equation}
    \tilde{c} = \frac{\qtr \left|\sqrt{\rho_\btheta} [L_1,L_2]\sqrt{\rho_\btheta}\right|}{2\sqrt{\mathcal{F}_{11}\mathcal{F}_{22}}},  \label{eq:ctilde}
\end{equation}
where $L_i$ are the SLD operators, $|A| = \sqrt{A^\dagger A}$, and $\qtr|A| = \qtrabs[A]$ for Hermitian (or anti-Hermitian) matrices. The IRTR is tight for pure states, i.e., there exists a measurement that attains equality in Eq.\ \eqref{eq:IRTR}. 

Lu and Wang further introduced a second form of their uncertainty relation in terms of the diagonal elements of the MSE. Using the single-parameter classical Cram\'er--Rao bound for each of the parameters, Eq.\ \eqref{eq:IRTR} can be recast to relate the diagonal elements of the MSE matrix 
\begin{equation}
    \gamma_1 +\gamma_2-2\sqrt{1-\tilde{c}^2}\sqrt{\left( 1-\gamma_1 \right)\left( 1-\gamma_2 \right)} \leq 2-\tilde{c}^2, \label{eq:LWUR}
\end{equation}
where $\gamma_j = 1/(V_{jj}\mathcal{F}_{jj})$. We refer to Eq.\ \eqref{eq:LWUR} as the Lu--Wang uncertainty relation (LWUR) and study its attainability by comparison to the NCRB. Central to this investigation is determining whether the NCRB and LWUR carry the same information about the attainable MSE, and thus, which estimation limit is practically more useful. 

We note that the IRTR and LWUR have been applied largely in situations where the incompatibility coefficient $\tilde{c}$ (Eq.\ \eqref{eq:ctilde}) takes the limiting values $0$ or $1$. In particular, it has been used to assess the trade-off for estimating the centroid and separation of incoherent optical point sources in configurations that lead to $\tilde{c}=0$ \cite{shi_joint_2023} and $\tilde{c}=1$ \cite{shao_performance-tradeoff_2022}. Ref.~\cite{xia_toward_2023} considered some intermediate values but with the goal of approaching the limit $\tilde{c}\rightarrow 0$. 

\section{Methods} \label{sec:methods}

We enable comparison between the Cram\'er--Rao bounds and the LWUR in two ways. This allows both the allowed precision (i.e., the smallest MSE-trace) and the uncertainty relation characteristics of the limits to be assessed and compared. 

The first method is to calculate a new estimation sum lower bound based on the LWUR, which we call the Lu--Wang Estimation Bound (LWB). It is the minimum sum of diagonal elements of the MSE allowed by the LWUR:
\begin{equation}
    \mathcal{C}_{\text{LW}} = \min_{v_1,v_2} \left\{ v_1+v_2 \ \big| \ \text{subject to Eq.\ \eqref{eq:LWUR}} \right\}, \label{eq:LWB}
\end{equation}
where $v_j$ is the mean squared error of the estimate of parameter $\theta_j$, i.e., $v_j = [V(\hbtheta)]_{jj}$. By definition, the LWB is a lower bound on the trace of the MSE. The minimisation problem is well-posed because it is convex, as shown in Appendix \ref{app:LWconvex}.

The second method is to determine the locus of MSE elements allowed by a Cram\'er--Rao bound with any weight matrix. That is, to convert a Cram\'er--Rao bound into an uncertainty relation that can be compared directly to the LWUR. This conversion was performed by Kull \textit{et al.} for some Cram\'er--Rao bounds \cite{kull_uncertainty_2020}. The uncertainty relation is developed by determining the intersection of the lines defined by the weighted Cram\'er--Rao bounds, as depicted in Figure \ref{fig:weightexample}\textbf{d}. For example, we may choose, for simplicity, a diagonal weight matrix $W = \operatorname{Diag}(w,2-w)$. The corresponding weighted Cram\'er--Rao bound, $\mathcal{C}(w)$, depends on the variable $w$ and the lower bound condition can be expressed as
\begin{equation}
    \ctr[WV] = w v_1+ (2-w)v_2 \geq \mathcal{C}(w). \label{eq:expandedlowerbound}
\end{equation}
The boundary of the overall region is then determined by maximising $v_2$ for a given $v_1$, or vice versa. 

\subsection{Qubit Nagaoka Uncertainty Relation} 
\label{sec:naguncert}
In the case of the NCRB for any qubit system, the above-described uncertainty relation construction can be performed explicitly \cite{funada_error_2020,funada_trade-off_2022,suzuki_parameter_2015}. Write the inverse SLDQFI as $\mathcal{F}^{-1} = \left(\begin{smallmatrix}
    a & c \\
    c & b 
\end{smallmatrix}\right)$ and the weight matrix as \mbox{$W = \operatorname{Diag}(w,2-w)$} with $0\leq w\leq 2$. Then, equality in Eq.~\eqref{eq:expandedlowerbound} with the NCRB in Eq.~\eqref{eq:NCRBanalytic} is
\begin{equation}
    w v_1+(2-w)v_2 = wa+(2-w)b+2\sqrt{w(2-w)}\sqrt{ab-c^2}. \label{eq:nagline}
\end{equation}
$v_2$ can then be maximised for fixed $v_1$ by simultaneously solving Eq.~\eqref{eq:nagline} and its derivative with respect to $w$ (see Appendix \ref{app:naguncertdetail} for details). The resultant relation can be expressed as
\begin{equation}
    \left(v_1-[\F^{-1}]_{11}\right)\left( v_2-[\F^{-1}]_{22} \right) \geq \frac{1}{\det[\F]}, \label{eq:naguncert}
\end{equation}
which defines a trade-off relation between the achievable precisions in simultaneously estimating two parameters. In particular, because the NCRB is tight for qubits, this trade-off relation is attainable, i.e., equality in Eq.~\eqref{eq:naguncert} can be achieved. In Sec.~\ref{sec:equal} and Sec.~\ref{sec:qubitrot} we compare this trade-off relation to the LWUR.

\section{Results}

In this section, we present our findings by comparing the NCRB, HCRB, LWB, and LWUR. We leave some calculations to the appendices. 

\subsection{Conditions for Equality (Qubits)} \label{sec:equal}
The main result in this subsection is the conditional equality of the NCRB and LWB for two-parameter estimation of qubit systems. This occurs when the incompatibility coefficient satisfies $\tilde{c}=1$.

First, we note that for a qubit system, the incompatibility coefficient satisfies $\tilde{c}=1$ if and only if the SLDFIM is diagonal. This follows from the following result (for qubits) in the Additional Note of Ref.\ \cite{nagaoka_new_2005},
\begin{equation}
    \qtrabs[\rho_\btheta [L_1,L_2]] = 2\sqrt{\det[\mathcal{F}]},
\end{equation}
and the definition of $\tilde{c}$, noting that $\qtrabs\left( \rho[X,Y] \right) = \qtrabs\left( \sqrt{\rho}[X,Y]\sqrt{\rho} \right) = \qtr|\sqrt{\rho}[X,Y]\sqrt{\rho}|$ \cite{horn_matrix_2017}. With \mbox{$\tilde{c}=1$}, the LWB simplifies to
\begin{equation}
    \mathcal{C}_{\text{LW, }\tilde{c}=1} = \min\left\{ v_1+v_2\ \bigg| \ \frac{1}{v_1\F_{11}}+ \frac{1}{v_2\F_{22}}\leq 1 \right\}. 
\end{equation}
This optimisation problem can be solved by the method of Lagrange multipliers to arrive at the result
\begin{equation}
    \mathcal{C}_{\text{LW, }\tilde{c}=1} = \left( \frac{1}{\sqrt{\F_{11}}}+\frac{1}{\sqrt{\F_{22}}} \right)^2, \label{eq:LWBc=1}
\end{equation}
see Appendix \ref{app:ceq1LWB} for details. As the SLDFIM is diagonal, this is the same as the NCRB given by Eq.\ \eqref{eq:NCRBanalytic}. That is, we have the following theorem:
\begin{thm}
    For two-parameter estimation of a qubit system, if $\tilde{c}=1$, then $\mathcal{C}_{\text{N}}=\mathcal{C}_{\text{LW}}$. 
    \label{thm}
\end{thm}
We note that this equality is a consequence of a stronger equivalence: for qubit systems with $\tilde{c}=1$, the LWUR is identical to the Nagaoka uncertainty relation determined in Sec.~\ref{sec:naguncert}.

In Sec.~\ref{sec:additionalnumerical}, we present numerical evidence that Theorem \ref{thm} may not be restricted to qubit systems.

\subsection{Examples of inequality} \label{sec:qubitrot}
In this subsection, we provide an example demonstrating that the NCRB and LWB need not be equal. To demonstrate this additional behaviour of the NCRB and LWB, we consider a model problem of estimating orthogonal rotations of a qubit state in the Bloch sphere. Consider a probe state with Bloch sphere coordinates $(r,\vartheta,\varphi)$, i.e.,
\begin{equation}
    \rho_0 = \frac{1}{2}\begin{pmatrix}
        1+r\cos\vartheta & re^{-i\varphi}\sin\vartheta \\
        re^{i\varphi}\sin\vartheta & 1-r\cos\vartheta.
    \end{pmatrix}.
\end{equation}
The parameters $\theta_x,\theta_y$ are the (small) magnitudes of rotations of the probe state about the $x$- and $y$-axes of the Bloch sphere, respectively, and are encoded via the unitary operator
\begin{equation}
    U(\theta_x,\theta_y) = e^{-i\theta_y\sigma_y/2}e^{-i\theta_x\sigma_x/2},
\end{equation}
where $\sigma_{x(y)}$ is the Pauli $X(Y)$ matrix. Accordingly, the derivatives of the encoded state with respect to the parameters are
\begin{equation}
    \frac{\partial \rho_\btheta}{\partial \theta_x}\bigg|_{\btheta=\mathbf{0}} = \frac{i}{2}[\rho_0,\sigma_x], \ \frac{\partial \rho_\btheta}{\partial \theta_y}\bigg|_{\btheta=\mathbf{0}} = \frac{i}{2}[\rho_0,\sigma_y].
\end{equation}
We find that the NCRB and HCRB depend on the probe state as 
\begin{equation}
    \mathcal{C}_{\text{N}} = \frac{(1+|\sec\vartheta|)^2}{r^2} \label{eq:rotNCRB}
\end{equation}
and 
\begin{equation}
    \mathcal{C}_{\text{H}} = \frac{1+2r|\sec\vartheta|+\sec^2\vartheta}{r^2}, \label{eq:rotHCRB}
\end{equation}
see Appendix \ref{app:rotNCRB} for details. Further, the incompatibility coefficient depends on $\vartheta$ and $\varphi$ as 
\begin{equation}
    \tilde{c} = \frac{|\cos\vartheta|}{\sqrt{\left( \cos^2\vartheta +\cos^2\varphi \sin^2\vartheta \right)\left( \cos^2\vartheta + \sin^2\varphi\sin^2\vartheta \right)}}, \label{eq:rotctilde}
\end{equation} 
which ranges from 0 to 1. Importantly, this allows us to assess any differences between the NCRB and LWB when the incompatibility coefficient is not equal to 1. In particular, here we present two examples: $\vartheta=\varphi=\pi/4$, and $\vartheta=\pi/2$.

\subsubsection{Example I: Finite Gap}
\begin{figure}
    \centering
    \includegraphics{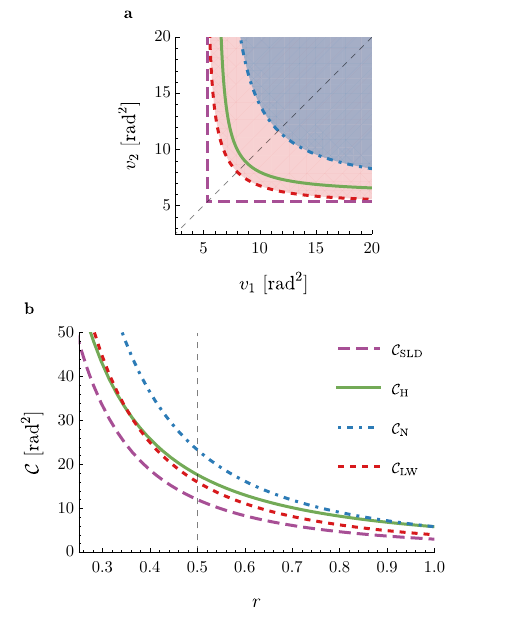}
    \caption{Estimation limits for qubit rotations with probe state $\vartheta=\varphi=\pi/4$. \textbf{a}, the regions defined by each Cram\'er--Rao bound and the LWUR when $r=1/2$. \textbf{b}, the Cram\'er--Rao bounds and the LWB as a function of $r$. The vertical dashed line in \textbf{b} indicates the value of $r$ at which the regions in \textbf{a} are drawn. }
    \label{fig:regionandbounds}
\end{figure}

For $\vartheta=\varphi=\pi/4$ and arbitrary $r\leq 1$, by Eq.\ \eqref{eq:rotNCRB} the NCRB is $(1+\sqrt{2})^2/r^2$, and by Eq.\ \eqref{eq:rotctilde} the incompatibility coefficient is $\tilde{c} = \sqrt{8/9}$. The LWB can then be computed analytically using the Karush--Kuhn--Tucker (KKT) conditions---see Appendix \ref{app:rotfingapLWB} for details---resulting in $\mathcal{C}_{\text{LW}} = 4/r^2$. In this example, the two lower bounds are unequal, as depicted in Figure \ref{fig:regionandbounds}, and the LWB is not attainable.

Furthermore, in this example for probe states with $r>1/\sqrt{8}$, the LWB is less informative than the HCRB (which has value $(3+2\sqrt{2}r)/r^2$). That is, the LWB cannot be saturated even with any collective measurement. 

\subsubsection{Example II: Infinite Gap}
As $\vartheta$ approaches $\pi/2$, the NCRB, HCRB, and SLDCRB diverge because the two orthogonal rotations are indistinguishable. Mathematically, the SLDQFI is singular. On the other hand, the incompatibility coefficient is zero, and the LWB can be computed exactly as
\begin{equation}
    \mathcal{C}_{\text{LW}} = \frac{1}{r^2\sin^2\varphi} + \frac{1}{r^2\cos^2\varphi}, \label{eq:pi/2LWB}
\end{equation}
which can be finite (see Appendix \ref{app:rotinfgapLWB} for details). As such, the gap between the NCRB and LWB can be infinitely large. This is an example where the LWUR fails to capture the inability to perform simultaneous estimation. In this particular case, the obstruction is not due to operator incompatibility; in the limit $\vartheta\rightarrow \pi/2$, the SLD operators commute. Instead, it is due to the degenerate encoding of the parameters which is only captured in the off-diagonal elements of the SLDQFI. 

\subsubsection{Further Comparison}
\begin{figure}
    \centering
    \includegraphics[width=\linewidth]{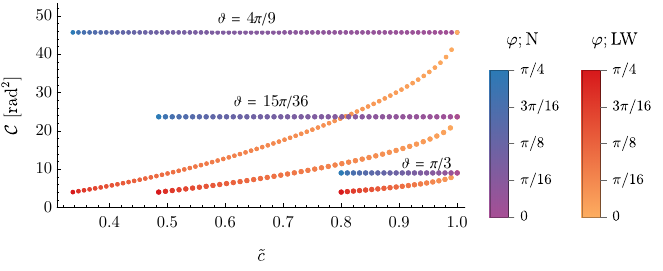}
    \caption{NCRB (blue-purple) and LWB (red-yellow) as a function of the incompatibility coefficient for the qubit rotations problem with a pure probe state. Series with varying $\varphi$ (full range $0\leq \varphi \leq \pi/4$) are shown for three values of $\vartheta$. From top to bottom, the series correspond to $\vartheta=4\pi/9$, $\vartheta=15\pi/36$, and $\vartheta=\pi/3$. That the NCRB is constant for each value of $\vartheta$ indicates that the ultimate attainable precision limit need not be related to the incompatibility coefficient.}
    \label{fig:vstildec}
\end{figure}

It is notable that for the qubit rotations problem, the NCRB (Eq.\ \eqref{eq:rotNCRB}) does not depend on $\varphi$ while the incompatibility coefficient (Eq.\ \eqref{eq:rotctilde}) does, which suggests that the LWB does too. This leads to a range of estimation problems where the NCRB and LWB are unequal. In Figure \ref{fig:vstildec} we present the NCRB and (numerically computed) LWB for three values of $\vartheta$. The bounds are plotted against the incompatibility coefficient, and the corresponding value of $\varphi$ is depicted in the colour. This demonstrates that the ultimate attainable precision limit (as the NCRB here is tight) need not be related to the incompatibility coefficient. We also see that the difference between the NCRB and LWB grows monotonically as the incompatibility coefficient tends away from~1.

\subsection{Additional numerical results} \label{sec:additionalnumerical}

\begin{figure*}
    \centering
    \includegraphics[width=\linewidth]{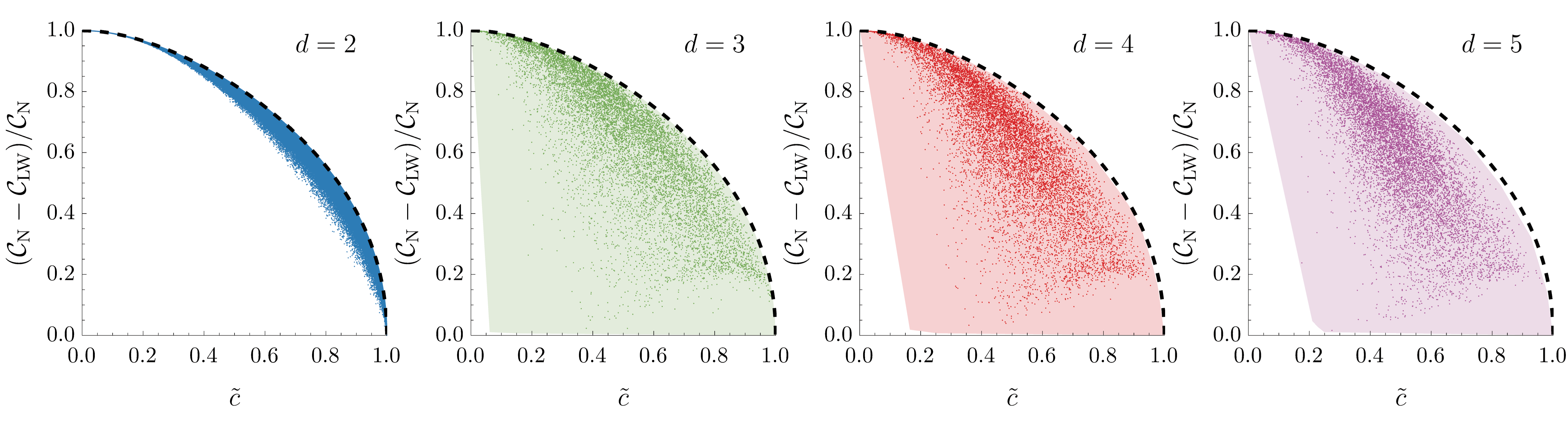}
    \caption{Normalised difference between the NCRB and LWB for random estimation problems in different Hilbert space dimensions $d$. The difference is plotted against the incompatibility coefficient $\tilde{c}$. The black dashed line in all subplots denotes the upper edge of points for $d=2$. All \num{1000000} points are plotted for $d=2$, but only \num{10000} out of \num{1000000} points are plotted for each $d\geq3$, so the convex hulls of all points are included as shaded regions marking where points were found.}
    \label{fig:randomprob}
\end{figure*}

To further assess the trends of the incompatibility coefficient and the difference between the NCRB and LWB, we extended our study to more general regimes, utilising random estimation problems in different Hilbert space dimensions. Calculating the estimation limits requires the probe state and its derivates with respect to the parameters. As such, a random density operator and two random traceless Hermitian operators constitute a random two-parameter estimation problem. We use density operators distributed normally in the Hilbert--Schmidt measure, see Ref.~\cite{johnston_qetlab_2016} for details.

In Figure \ref{fig:randomprob}, we present the normalised difference between the respective NCRB and LWB obtained numerically against the corresponding incompatibility coefficients, separated by dimension. We also include a point where the NCRB and LWB are equal and $\tilde{c}=1$, as this is possible as we show in Appendix \ref{app:equalanydim}. This point extends the shaded regions to the lower right corner. For dimension $d=2$, the results align with the findings from the qubit rotations problem: the gap is zero when $\tilde{c}=1$ and the normalised gap tends to one as $\tilde{c}$ tends to zero, indicating a large disagreement between the bounds. For higher dimensions, the points are bounded above by a monotonic function of $\tilde{c}$ that is zero at $\tilde{c}=1$. This suggests that the ordering $\mathcal{C}_{\text{LW}}\leq \mathcal{C}_{\text{N}}$ holds, and that equality holds whenever $\tilde{c}=1$.

\subsection{Diagreement between the IRTR, LWUR, and NCRB} \label{sec:discrepancy}
In the previous subsections, we have seen that the LWB can be unattainable even for estimating pure states. This raises the question of how such a discrepancy arises, given that the IRTR (Eq.\ \eqref{eq:IRTR}) is attainable for pure states. Here, we demonstrate that this discrepancy can be attributed to the differing figure-of-merit for estimation performance, and that an IRTR-optimal measurement (i.e., one saturating Eq.~\eqref{eq:IRTR}) need not provide an optimal MSE. We define the IRTR-gap of a measurement to be the quantity
\begin{equation}
    G \coloneq \Delta_1^2+\Delta_2^2 + 2\sqrt{1-\tilde{c}^2}\Delta_1\Delta_2 - \tilde{c}^2,
\end{equation}
which is non-negative by Eq.\ \eqref{eq:IRTR}. A value of $G=0$ is necessary for the saturation of the LWUR.  

\begin{figure}
    \centering
    \includegraphics[width=0.85\linewidth]{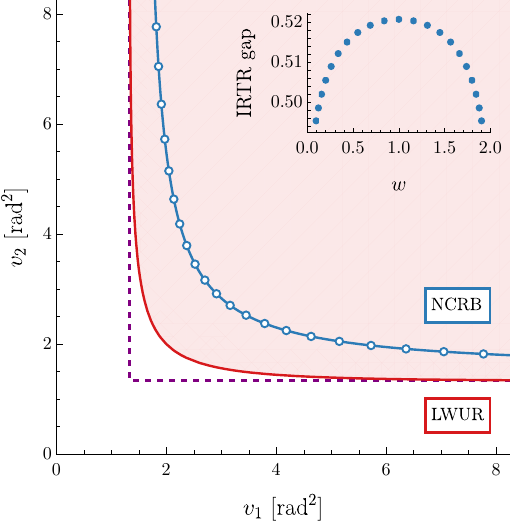}
    \caption{Difference between the NCRB and LWUR, and the IRTR gap, for the qubit rotations problem with $(r,\vartheta,\varphi)=(1,\pi/4,\pi/4)$. Main: MSE elements $(v_1,v_2)$ of measurements saturating the weighted NCRB, but which do not saturate the LWUR (as it is unattainable). The measurements are optimised for weight matrices of the form $\operatorname{Diag}(w,2-w)$. The dashed line indicates the boundary of the region defined by the SLDCRB. Inset: non-zero gap of the IRTR for the measurements in the main figure.}
    \label{fig:regiongap}
\end{figure}

For demonstration, we again consider the qubit rotations problem from Sec.\ \ref{sec:qubitrot} with $\vartheta=\varphi=\pi/4$ and $r=1$. First, we find that for this problem, the LWB being unattainable is a particular case of the LWUR being unattainable---the measurements that saturate the NCRB (and are thus optimal in the sense of minimising the total MSE) for different weight matrices do not saturate the LWUR, and thus no measurement can saturate the LWUR. This behaviour is presented in Figure \ref{fig:regiongap}, where we also present the non-zero gap of the IRTR for the same weighted MSE-optimal measurements. 

\begin{figure}
    \centering
    \includegraphics{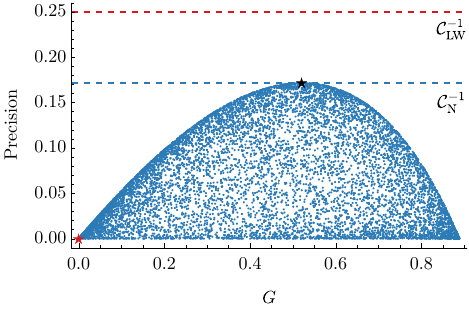}
    \caption{Estimation figures of merit of random rank-1 measurements for the $(r,\vartheta,\varphi) = (1,\pi/4,\pi/4)$ qubit rotations problem. The vertical axis is the inverse of the trace of the MSE matrix, calculated as $1/\ctr[F^{-1}]$ in units of $\mathrm{rad}^{-2}$, which we call the measurement ``precision''. The fact that the maximum vertical value does not occur at $G=0$ indicates that the MSE (here calculated via the classical Fisher information) and the IRTR disagree on the optimal measurement. The black star denotes the most informative measurement (minimising MSE and saturating the NCRB), and the red star denotes a measurement saturating the IRTR. The minimum value of $G$ from \num{10000} random measurements was less than $2\times 10^{-7}$.}
    \label{fig:randmeas}
\end{figure}

We use random measurements to investigate the behaviour of the MSE and IRTR-gap in this estimation problem. Measurements were constructed by generating three random rank-1 operators $\ketbra{\psi_i}{\psi_i}$ and normalising as in Refs.\ \cite{johnston_qetlab_2016,heinosaari_random_2020}. Two figures of merit for estimation performance were calculated for each measurement, and we present the results in Figure \ref{fig:randmeas}. Here, for the MSE figure of merit, we plot the inverse of the trace of the MSE matrix, which we call the measurement ``precision''. The MSE matrix is calculated as the inverse of the classical Fisher information matrix because the classical Cram\'er--Rao bound can be saturated with an appropriate estimator. 

In this problem, the MSE is not minimised when the IRTR-gap is zero. Furthermore, as the IRTR-gap tends to zero, the MSE diverges---the IRTR-optimal measurement does not gain any information about the two parameters. This means that the information regret, at least how it is implemented in the IRTR, is not always a practical metric for quantifying parameter estimation performance. Equivalent plots for different probe states are included in Appendix \ref{app:extraplots}, which demonstrate different positions of the peak. In Appendix \ref{app:extraplots}, we also present results for measurements described by full-rank POVMs, which do not perform better than rank-1 POVMs in either figure of merit. 

We note that the divergence of the MSE as the IRTR-gap tends to zero is a consequence of the classical Fisher information matrix becoming singular as $G$ decreases. One measurement that saturates the IRTR ($G=0$) is the two-outcome POVM with elements $\Pi_1 = \ketbra{0}{0}$, $\Pi_2 = \ketbra{1}{1}$ which has the singular classical Fisher information matrix $F = \left(\begin{smallmatrix}
    1/2 & -1/2\\
    -1/2 & 1/2
\end{smallmatrix}\right)$. While such a measurement cannot be paired with an unbiased estimator, it supports the finding that the MSE and information regret figures of merit do not in general agree.

\section{Discussion and Conclusion}

We report that there are estimation applications for which the NCRB and LWB are equal, and thus that the NCRB and LWUR set the same limit on the attainable mean squared error. Indeed, this is the case for qubits whenever the incompatibility coefficient satisfies $\tilde{c}=1$, or equivalently, the SLDFIM is diagonal. The numerical study of random estimation problems for higher dimensions suggests that the NCRB and LWB are equal whenever $\tilde{c}=1$, irrespective of the Hilbert space dimension. This has potential utility because, in such cases, the LWB has an exact expression (Eq.\ \eqref{eq:LWBc=1}). 

On the other hand, there are scenarios where the gap between the NCRB and LWB can be arbitrarily large. The LWUR is thus not in general an attainable precision limit. Interestingly, we also find (see Figure \ref{fig:vstildec}) that the best attainable precision (here, the NCRB) need not be related to the incompatibility coefficient, $\tilde{c}$. This suggests that $\tilde{c}$ cannot reliably be used as an indicator of compatible parameters; a decreased value of $\tilde{c}$ need not mean that the chosen probe state mitigates the error trade-off. In particular, the condition $\tilde{c}=0$ is not sufficient for the SLDCRB to be saturated and is not necessarily equivalent to the optimal observables being compatible. The lack of attainability of the LWUR can be present even when the probe state is pure, even though the Lu--Wang IRTR is tight for pure states. 

In particular, we investigated the estimation of orthogonal rotations of a qubit in the Bloch sphere, showcasing the discrepancy between the NCRB and LWB. In the configuration of $(r,\vartheta,\varphi) = (1,\pi/4,\pi/4)$, we determined that the IRTR can be saturated, but the classical Fisher information matrix of the corresponding saturating measurements is singular. As such, the IRTR-optimal measurement gains no information about the magnitudes of the orthogonal rotations, while the MSE-optimal measurement does not saturate the IRTR. We note that the classical Fisher information matrix elements of the IRTR-optimal measurement are finite and non-zero. As such, it appears that the lack of consideration for the off-diagonal elements of the information regret matrix in the IRTR is detrimental to its general performance.

We find that the LWUR does not provide any information that is not already contained in the NCRB. For qubit estimation problems, this is unsurprising because the NCRB is tight. It is possible that for cases where we do not have a tight bound, for example when $d>2$ or for more than two parameters, some uncertainty relation may give rise to tighter bounds for metrology. We leave this as an open question. 

\bibliography{bib}

\section*{Acknowledgements}
This research was funded by the Australian Research Council Centre of Excellence CE170100012. This research was also supported by A*STAR C230917010, Emerging Technology and A*STAR C230917004, Quantum Sensing. S.Y. was supported by the ANU Dunbar Physics Honours Scholarship and the A*STAR Singapore International Pre-Graduate Award. 

\appendix

\section{Convexity of Lu--Wang Bound Minimisation Problem} \label{app:LWconvex}

The LWB minimisation problem is convex because (i) its objective ($v_1+v_2$) is convex and (ii) its feasible region is convex. Condition (ii) can be proven as follows. The constraint function in Eq.~\eqref{eq:LWB} can be written as 
\begin{multline}
    \underbrace{\frac{1}{\alpha v_1}+\frac{1}{\beta v_2}}_{g_1}+\tilde{c}^2-2 \\-2\sqrt{1-\tilde{c}^2}\underbrace{\sqrt{\left( 1-\frac{1}{\alpha v_1} \right)\left( 1-\frac{1}{\beta v_2} \right)}}_{-g_2}\leq 0,
\end{multline}
for SLDQFI of the form $\F = \left(\begin{smallmatrix}
\alpha & \delta \\
\delta & \beta
\end{smallmatrix}\right)$, along with the condition that $\alpha v_1 \geq 1$ and $\beta v_2\geq 1$ to satisfy the single-parameter SLDCRB. 
By linearity, the left-hand side is convex if the functions $g_1$, $g_2$ are convex. The Hessian of $g_1$ is positive semidefinite:
\begin{equation}
    H_{g_1} = \begin{pmatrix}
        \frac{\partial^2 g_1}{\partial v_1^2} & \frac{\partial^2 g_1}{\partial v_1\partial v_2} \\
        \frac{\partial^2 g_1}{\partial v_2 \partial v_1} & \frac{\partial^2 g_1}{\partial v_2^2}
    \end{pmatrix} = \begin{pmatrix}
        \frac{2}{\alpha v_1^3} & 0 \\
        0 & \frac{2}{\beta v_2^3}
    \end{pmatrix}.
\end{equation}
The Hessian of $g_2$ has elements

\begin{align}
    \begin{split}
        \left[ H_{g_2} \right]_{11} &= \frac{\left( 4\alpha v_1-3 \right)\sqrt{\left( 1-\frac{1}{\alpha v_1} \right)\left( 1-\frac{1}{\beta v_2} \right)}}{4v_1^2(\alpha v_1-1)^2}, \\
        \left[ H_{g_2} \right]_{12} &=\frac{-1}{4v_1^2v_2^2 \alpha \beta \sqrt{\left( 1-\frac{1}{\alpha v_1} \right)\left( 1-\frac{1}{\beta v_2} \right)}}, \\
        \left[ H_{g_2} \right]_{22} &= \frac{\left( 4\beta v_2-3 \right)\sqrt{\left( 1-\frac{1}{\alpha v_1} \right)\left( 1-\frac{1}{\beta v_2} \right)}}{4v_2^2(\beta v_2-1)^2}.
    \end{split}
\end{align}
The matrix can then be shown to be positive semidefinite because $z^\top H_{g_2} z\geq 0$ for all non-zero real vectors $z = (z_1,z_2)^\top$, which can be expressed as 
\begin{equation}
    z^\top H_{g_2} z = \frac{r_1^2 +r_2^2 +\left( \frac{r_1}{2\sqrt{v_1\alpha -1}}-\frac{r_2}{2\sqrt{v_2\beta-1}} \right)^2}{\sqrt{v_1v_2\alpha\beta \left( v_1\alpha-1 \right)\left( v_2\beta-1 \right)}},
\end{equation}
where 
\begin{equation}
    r_1 = \frac{z_1}{v_1}\sqrt{v_2\beta-1}, \quad r_2 = \frac{z_2}{v_2}\sqrt{v_1\alpha -1}.
\end{equation}
Hence, $z^\top H_{g_2}z$ is non-negative and $H_{g_2}$ is positive semidefinite. It thus follows that the LWB minimisation problem is convex. 

Furthermore, the interior of the feasible region is non-empty, so strong duality holds. 

\section{Qubit Nagaoka Uncertainty Relation Calculation} \label{app:naguncertdetail}

Eq.~\eqref{eq:nagline} can be rearranged as
\begin{equation}
	v_2 = \frac{w}{2-w}(a-v_1) + b + 2\sqrt{\frac{w}{2-w}}\sqrt{ab-c^2}. \label{eq:b1}
\end{equation}
Keeping $v_1$ fixed, coordinates $(v_1,v_2)$ on the uncertainty relation curve can be determined by maximising $v_2$ with respect to $w$ by finding stationary points:
\begin{multline}
	0 = \frac{\partial v_2}{\partial w} = (a-v_1)\left(\frac{1}{2-w} + \frac{w}{(2-w)^2}\right) \\+ \sqrt{\frac{2-w}{w}}\frac{2}{(2-w)^2} \sqrt{ab-c^2}. \label{eq:b2}
\end{multline}
Multiplying both sides by $(2-w)$ gives
\begin{align}
	0 &=(a-v_1)\left(1 + \frac{w}{2-w}\right) + \sqrt{\frac{2-w}{w}}\frac{2}{2-w}\sqrt{ab-c^2} \\
	&= (a-v_1)\frac{2}{2-w}+\sqrt{\frac{2-w}{w}}\frac{2}{2-w}\sqrt{ab-c^2}.
\end{align}
Then, multiplying by $(2-w)/2$ and rearranging gives
\begin{equation}
	\sqrt{\frac{2-w}{w}} = \frac{v_1-a}{\sqrt{ab-c^2}}. \label{eq:B5}
\end{equation}
Eq.~\eqref{eq:B5} can be substituted into Eq.~\eqref{eq:b1} to obtain
\begin{equation}
	v_2 = \frac{ab-c^2}{(v_1-a)^2}(a-v_1) + b + \frac{2(ab-c^2)}{v_1-a}. 
\end{equation}
Rearranging yields Eq.~\eqref{eq:naguncert},
\begin{equation}
	(v_1-a)(v_2-b) = -(ab-c^2) + 2(ab-c^2)= ab-c^2. 
\end{equation}

\section{$\tilde{c}=1$ LWB Computation} \label{app:ceq1LWB}

For $\tilde{c}=1$, the LWB minimisation problem for SLDQFI $\F = \left( \begin{smallmatrix}
    \F_{11} & \F_{12} \\
    \F_{21} & \F_{22}
\end{smallmatrix} \right)$ is
\begin{equation}
    \min\left\{ f_0=v_1+v_2 \ \bigg|\ f_1 = \frac{1}{v_1\F_{11}}+\frac{1}{v_2\F_{22}}-1\leq 0 \right\}.
\end{equation}
Because strong duality holds, we can determine the solution by solving the Lagrange dual problem. The Lagrange dual function is 
\begin{align}
    \begin{split}
        g(\lambda) &= \inf_{v_1,v_2}\left\{ f_0(v_1,v_2)+\lambda f_1(v_1,v_2) \right\} \\
        &= \inf_{v_1,v_2} \left\{ v_1 + v_2 + \lambda \left( \frac{1}{v_1\F_{11}} + \frac{1}{v_2\F_{22}}-1 \right) \right\}\\
        &= \inf_{v_1}\left\{ v_1+\frac{\lambda}{v_1\F_{11}} \right\}+\inf_{v_2}\left\{ v_2+\frac{\lambda}{v_2\F_{22}} \right\}-\lambda.
    \end{split}
\end{align}
Each infimum can be solved by finding the stationary point, resulting in
\begin{equation}
    g(\lambda) = \sqrt{\lambda}\left( \frac{2}{\sqrt{\F_{11}}}+\frac{2}{\sqrt{\F_{22}}} \right)-\lambda.
\end{equation}
Finally, the LWB is equal to the supremum of the Lagrangian, subject to $\lambda\geq 0$:
\begin{equation}
    \mathcal{C}_{\text{LW}} = \sup_\lambda\left\{ g(\lambda) \ | \ \lambda \geq 0 \right\} = \left( \frac{1}{\sqrt{\F_{11}}}+\frac{1}{\sqrt{\F_{22}}} \right)^2.
\end{equation}

\section{SLDQFI, NCRB, and HCRB Computation} \label{app:rotNCRB}

The SLD operators can be computed using the eigenbasis of the density operator and the derivatives of the density operator \cite{paris_quantum_2009}. The operators are
\begin{align}
    L_x &= \begin{pmatrix}
        r\sin\varphi\sin\vartheta & ir\cos\vartheta \\
        -ir\cos\vartheta & -r\sin\varphi\sin\vartheta
    \end{pmatrix},\\
    L_y &= \begin{pmatrix}
        -r\cos\varphi\sin\vartheta & r\cos\varphi \\
        r\cos\vartheta & r\cos\varphi\sin\vartheta 
    \end{pmatrix}. 
\end{align}
The SLDQFI is then computed via Eq.~\eqref{eq:sldqfi} as
\begin{equation}
    \mathcal{F} = r^2\begin{pmatrix}
        1-\cos^2\varphi\sin^2\vartheta & -\sin\varphi\cos\varphi\sin^2\vartheta \\
        -\sin\varphi\cos\varphi\sin^2\vartheta & 1-\sin^2\varphi\sin^2\vartheta
    \end{pmatrix}.
\end{equation}
The NCRB can then be computed by Eq.~\eqref{eq:NCRBanalytic} to get the result in Eq.~\eqref{eq:rotNCRB}. 

The exact expression for the HCRB relies on the RLD operators, which can be computed directly by inverting the density matrix: $L_j^R = \rho_{\boldsymbol{\theta}}^{-1}\partial_j\rho_{\boldsymbol{\theta}}$. The RLD Fisher information matrix (with elements $\qtr\left[\rho_{\boldsymbol{\theta}} L_j^R(L_i^R)^\dagger \right]$) is:
\begin{equation}
    \mathcal{F}_R = \frac{r^2}{1-r^2}\begin{pmatrix}
        1-\cos^2\varphi\sin^2\vartheta & \alpha \\
        \alpha^* & 1-\sin^2\varphi\sin^2\vartheta
    \end{pmatrix},
\end{equation}
with $\alpha = i r \cos\theta-\cos\varphi\sin\varphi\sin^2\vartheta$. The HCRB can then be computed directly as \cite{suzuki_explicit_2016}
\begin{equation}
    \mathcal{C}_{\text{H}}  = \ctr[\re[\F_R^{-1}]]+ \ctrabs[\operatorname{Im}[\F_R^{-1}]],
\end{equation}
resulting in Eq.~\eqref{eq:rotHCRB}.

\section{Finite Gap LWB Computation} \label{app:rotfingapLWB}

With $\vartheta=\varphi=\pi/4$ we have $\tilde{c}=\sqrt{8/9}$ and SLDQFI $\F = r^2\left( \begin{smallmatrix}
    \frac{3}{4} & -\frac{1}{4} \\
    -\frac{1}{4} & \frac{3}{4}
\end{smallmatrix} \right)$. For simplicity, we define $\kappa = 1/\F_{11} = 1/\F_{22} = 4/(3r^2)$. We use the Karush--Kuhn--Tucker (KKT) contidions \cite{boyd_convex_2004} to solve the minimisation problem, which can be stated as 
\begin{multline}
    \mathcal{C}_{\text{LW}} = \min\Bigg\{ v_1+v_2 \ \bigg| \ f_1(v_1,v_2) = \frac{\kappa}{v_1} + \frac{\kappa}{v_2} \\-\frac{2}{3}\sqrt{\left( 1-\frac{\kappa}{v_1} \right)\left( 1-\frac{\kappa}{v_2} \right)}-\frac{10}{9}\leq 0, \\
    f_2(v_1,v_2) = \kappa-v_1 \leq 0, \ f_3(v_1,v_2) = \kappa-v_2\leq 0  \Bigg\}.
\end{multline}

The relevant KKT conditions are as follows, where `$\star$' denotes optimisers. The conditions are necessary and sufficient for the solution of the convex problem because strong duality holds.  
\begin{equation}
    \begin{aligned}
        &\text{(I)} &\quad f_1(v_1^\star,v_2^\star)\leq 0, \ \kappa-v_1^\star \leq 0, \ \kappa-v_2^\star \leq 0;\\
        &\text{(II)} &\quad \lambda_1^\star \geq 0, \ \lambda_2^\star \geq 0, \ \lambda_3^\star \geq 0, \\
        &\text{(III)} & \lambda_1^\star f_1(v_1^\star,v_2^\star)=0, \ \lambda_2^\star(\kappa-v_1^\star)=0, \ \lambda_3^\star(\kappa-v_2^\star)=0, \\
        &\text{(IV)} & \grad f_0(v_1^\star,v_2^\star) + \lambda_1^\star \grad f_1(v_1^\star,v_2^\star) + \lambda_2^\star \grad (\kappa-v_1^\star)\\
        & & + \lambda_3^\star \grad (\kappa-v_2^\star) =0.
    \end{aligned}
\end{equation}
The complementary slackness conditions (III) can have three options: (i) both $\lambda_2^\star=0$ and $\lambda_3^\star=0$, which leads to a contradiction with (I). (ii) one of $\lambda_2^\star$, $\lambda_3^\star$ is zero, which upon enforcing $f_1(v_1,v_2)\leq0$ from (I) leads to $v_1^\star+v_2^\star\geq 10\kappa$. The better option is, (iii) $\lambda_2^\star=\lambda_3^\star=0$. In this case, condition (IV) can be written as (dropping `$\star$' for brevity) 
\begin{equation}
    \begin{pmatrix}
        1-\frac{\lambda_1\kappa}{v_1^2}\left( 1+\frac{1}{3} \sqrt{\frac{1-\frac{\kappa}{v_2}}{1-\frac{\kappa}{v_1}}} \right)\\
        1-\frac{\lambda_1\kappa}{v_2^2}\left( 1+\frac{1}{3} \sqrt{\frac{1-\frac{\kappa}{v_1}}{1-\frac{\kappa}{v_2}}} \right)
    \end{pmatrix} = \begin{pmatrix}
        0 \\
        0
    \end{pmatrix} \label{eq:D3}
\end{equation}
which implies
\begin{equation}
    \frac{v_1^2}{1+\frac{1}{3} \sqrt{\frac{1-\frac{\kappa}{v_2}}{1-\frac{\kappa}{v_1}}}} = \frac{v_2^2}{1+\frac{1}{3} \sqrt{\frac{1-\frac{\kappa}{v_1}}{1-\frac{\kappa}{v_2}}}}.
\end{equation}
This has a solution of $v_1=v_2\equiv v$ (recall that the KKT conditions are sufficient). Finally, as $\lambda_1^\star$ is non-zero (see Eq.~\eqref{eq:D3}), from (III) we must have $f_1(v,v) = 0$. Solving this, we find
\begin{equation}
    v_1^\star = v_2^\star = v = \frac{3\kappa}{2} = \frac{2}{r^2},
\end{equation}
which gives the LWB as $\mathcal{C}_{\text{LW}} = 4/r^2$. 

\section{Infinite Gap LWB Computation} \label{app:rotinfgapLWB}

For $\tilde{c}=0$, the LWUR can be written as 
\begin{equation}
    \frac{1}{v_1\F_{11}}+\frac{1}{v_2\F_{22}} -2\sqrt{\left( 1-\frac{1}{v_1\F_{11}} \right)\left( 1-\frac{1}{v_2\F_{22}} \right)} \leq 2,
\end{equation}
which can be rewritten as 
\begin{equation}
    \left( \sqrt{1-\frac{1}{v_1\F_{11}}} + \sqrt{1-\frac{1}{v_2\F_{22}}} \right)^2 \geq 0.\label{eq:E2}
\end{equation}
This implies that the term inside the brackets in Eq.~\eqref{eq:E2} is real-valued. As $\sqrt{ 1-\frac{1}{v_i\F_{ii}}}$ is real-valued, we must have that 
\begin{equation}
    1-\frac{1}{v_1\F_{11}} \geq 0 \ \text{and} \ 1-\frac{1}{v_2\F_{22}} \geq 0,
\end{equation}
which are the single-parameter SLDCRB conditions. That is, when $\tilde{c}=0$, the LWUR imposes only the single-parameter SLDCRB for each parameter. 

Accordingly, the LWB for $\tilde{c}=0$ can be simply computed as 
\begin{equation}
    \mathcal{C}_{\text{LW, }\tilde{c}=0} = \frac{1}{\F_{11}} + \frac{1}{\F_{22}} = \frac{\F_{11} +\F_{22}}{\F_{11}\F_{22}}.
\end{equation}
We note that this is bounded above by the SLDCRB, which is 
\begin{equation}
    \mathcal{C}_{\text{SLD}} = \frac{\F_{11}+\F_{22}}{\F_{11}\F_{22}-\F_{12}^2}.
\end{equation}

For the qubit rotations problem with $\vartheta=\pi/2$, the SLDQFI is 
\begin{equation}
    \F = r^2\begin{pmatrix}
        \sin^2\varphi & -\sin\varphi\cos\varphi \\
        -\sin\varphi\cos\varphi & \cos^2\varphi,
    \end{pmatrix}
\end{equation}
from which the LWB in Eq.~\eqref{eq:pi/2LWB} can be computed. 

\section{Example where NCRB and LWB are equal in any dimension}\label{app:equalanydim}

Here, we provide an estimation example in any finite-dimensional Hilbert space where $\tilde{c}=1$ and the NCRB and LWB are equal. Consider the $d$-dimensional probe state
\begin{equation}
    \rho = \begin{pmatrix}
        \frac{1}{d-\frac{1}{2}} & & & \\
        & \ddots & & \\
        & & \frac{1}{d-\frac{1}{2}} & \\
        & & & 1-\frac{d-1}{d-\frac{1}{2}}
    \end{pmatrix}
\end{equation}
and derivatives $\partial_i\rho = \frac{i}{2}[\rho,M_i]$ where 
\begin{equation}
    M_1 = \begin{pmatrix}
        & & & 1 \\
        & & &  \\
        & & & \\
        1 & & & 
    \end{pmatrix}, \ M_2 = \begin{pmatrix}
        & & & -i \\
        & & & \\
        & & & \\
        i & & & 
    \end{pmatrix},
\end{equation}
where blank entries are zero. It is straightforward to calculate that the SLD operators are $L_1 = -M_2/3$ and $L_2 = M_1/3$, that the SLDQFI is 
\begin{equation}
    \F = \begin{pmatrix}
        \frac{1}{6d-3} & 0 \\
        0 & \frac{1}{6d-3}
    \end{pmatrix},
\end{equation}
and that $\tilde{c}=1$. Hence, the LWB is given by Eq.~\eqref{eq:LWBc=1} and is 
\begin{equation}
    \mathcal{C}_{\text{LW}} = 12(2d-1). 
\end{equation}
For the NCRB, an ansatz for the $X$ matrices in Eq.~\eqref{eq:nagfull} is 
\begin{equation}
    X_i = \begin{pmatrix}
        & & & a_i \\
        & & &  \\
        & & & \\
        b_i & & & 
    \end{pmatrix}.
\end{equation}
The locally unbiased conditions enforce that $a_1=  i(2d-1)$, $b_1 = -i(2d-1)$, $a_2 = 2d-1$, $b_2 = 2d-1$, which results in an upper bound to the NCRB (as the ansatz is not necessarily optimal) of
\begin{equation}
    \mathcal{C}_{\text{N}} \leq \ctr[Z(X)] + \qtrabs[\rho[X_1,X_2]] = 12(2d-1). \label{eq:F6}
\end{equation}
By comparing to the numerical solution of the semidefinite program, we find the $X$ ansatz is optimal and Eq.~\eqref{eq:F6} has equality. Thus, for this problem, $\tilde{c}=1$ and the NCRB and LWB are equal. 

\begin{figure*}
    \centering
    \includegraphics{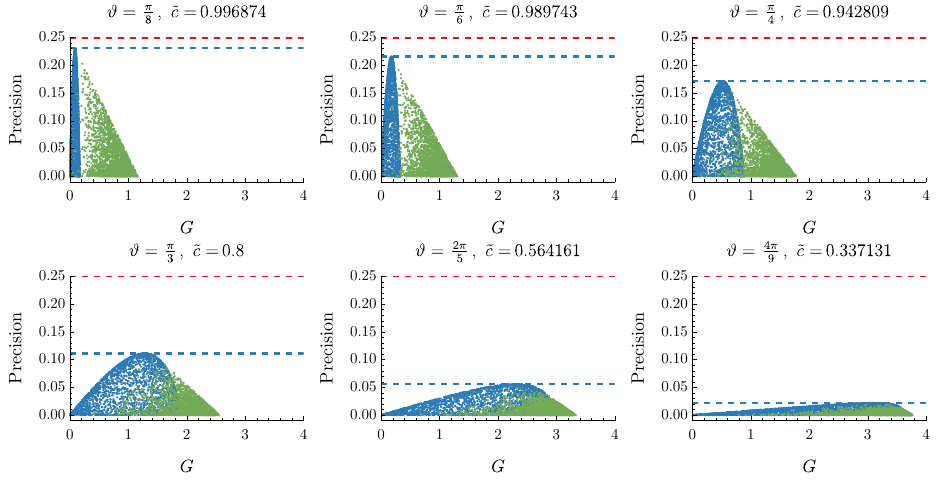}
    \caption{Estimation figures of merit for random measurements of different configurations of the qubit rotations problem, as in Fig.~\ref{fig:randmeas}. The probe states have $r=1$, $\varphi=\pi/4$, and $\vartheta$ as denoted on each plot. The blue points are for measurements with rank-1 POVMs, and the green points are for non-rank-1 POVMs (i.e., rank-2). The blue dashed line denotes the inverse of the NCRB and the red dashed line denotes the inverse of the LWB.}
    \label{fig:randommeasseries}
\end{figure*}

\section{Further comparison of figures of merit} \label{app:extraplots}

In Figure \ref{fig:randommeasseries}, we present the results of random measurements, as in Figure \ref{fig:randmeas}, for different probe states. Here, we also include the results for rank-2 POVMs, which perform worse than the rank-1 POVMs. Curiously, the envelope of the rank-2 POVMs, if extended, appears to coincide with the LWB at $G=0$.

\end{document}